\documentclass[a4paper,twocolumn]{emulateapj}
\usepackage{amstext}
\usepackage{amsmath}
\usepackage{amssymb}
\usepackage[colorlinks=true,linkcolor=blue,citecolor=blue]{hyperref}
\begin{document}

\slugcomment{Draft \today}
\slugcomment{Submitted to ApJL}

\shorttitle{A Stellar Wind Origin for the G2 Cloud: Numerical Simulations}
\shortauthors{De Colle et al.}


\title{A STELLAR WIND ORIGIN FOR THE G2 CLOUD: THREE-DIMENSIONAL NUMERICAL SIMULATIONS}

\author{Fabio De Colle, A.C. Raga, Flavio F. Contreras-Torres, Juan C. Toledo-Roy\altaffilmark{1}}
\altaffiltext{1}{Instituto de Ciencias Nucleares, Universidad Nacional Aut{\'o}noma de M{\'e}xico, A. P. 70-543 04510 D. F. Mexico}
\email{fabio@nucleares.unam.mx}




\begin{abstract}
We present 3D, adaptive mesh refinement simulations
of G2, a cloud of gas moving in a highly eccentric orbit towards the galactic center. 
We assume that G2 originates from a stellar wind interacting with the environment of the Sgr A* black hole.
The stellar wind forms a cometary bubble which becomes increasingly elongated 
as the star approaches periastron. A few months after periastron passage, 
streams of material begin to accrete on the central black
hole with accretion rates $\dot{M} \sim 10^{-8}$~M$_\sun$ yr$^{-1}$.
Predicted Br$\gamma$ emission maps and position-velocity diagrams show 
an elongated emission resembling recent observations of G2.
A large increase in luminosity is predicted by the emission coming from the 
shocked wind region during periastron passage. The observations, showing 
a constant Br$\gamma$ luminosity, remain puzzling, and are explained here
assuming that the emission is dominated by the free-wind region.
The observed Br$\gamma$ luminosity ($\sim 8 \times 10^{30}$~erg s$^{-1}$) is
reproduced by a model with a $v_w=50$~km s$^{-1}$ wind velocity and a 
$10^{-7}$~M$_\sun$ yr$^{-1}$ mass loss rate if the emission comes from the
shocked wind. A faster and less dense wind reproduces the Br$\gamma$ luminosity 
if the emission comes from the inner, free wind region.
The extended cometary wind bubble, largely destroyed by the tidal interaction 
with the black hole, reforms a few years after periastron passage.
As a result, the Br$\gamma$ emission is more compact after periastron passage.
\end{abstract}

  



\section{Introduction}
\label{sec1}

The G2 cloud is falling towards the Sgr A* black hole (BH hereafter), and was first
detected in Br$\gamma$ images by \cite{2012Natur.481...51G}, with
an estimated mass of $\sim 3$ Earth masses. Its
closest approach is estimated to occur in early 2014
\citep{2013MNRAS.435L..19D}, or mid-March 2014
\citep{2013ApJ...773L..13P},
at a distance of $\sim 100$~AU from the BH. 

\citet{2013ApJ...744...44G,2013ApJ...763...78G} show that the cloud has developed an
elongated structure. Part of the material might have already
gone beyond periastron, as evidenced by the large velocity jump
observed in PV diagrams of the extended Br$\gamma$ emission.
This stretching of the cloud (due to the tidal forces of
the central BH) will result in an extended cloud/BH
interaction, lasting for $\sim 1$ yr.

Even though the region around the galactic center has been extensively monitored
in radio wavelengths (see, e.g., \citealt{2013ATel.5159....1B, 2014Atel.5727.1C}) and in
X-rays (see \citealt{2014Atel.5861.1C}), no emission from the G2 cloud has
yet been detected. This lack of detection at radio and X-ray
wavelengths provides an interesting constraint on theoretical models
of the G2 cloud
 \citep{2012ApJ...757L..20N, 2013MNRAS.436.1955C, 2013MNRAS.432..478S,
2013ApJ...770L..21Y,2013PhRvL.110v1102B,2013MNRAS.435L..19D}.

Two possible scenarios for the G2 cloud have been studied:
\begin{itemize}
\item[$a.$] that it is an isolated, ISM cloud with an initial spherical
\citep{2012Natur.481...51G}
or shell-like \citep{2012AA...546L...2M,2012ApJ...750...58B,
2012ApJ...755..155S} density distribution,
\item[$b.$] that it is a wind bubble \citep{scoville}
or a photo-evaporated proto-planetary disk \citep{murray,2012ApJ...756...86M} 
and therefore has a central (undetected) stellar nucleus.
\end{itemize}
Scenario $a$ has been studied with 2D and 3D hydrodynamic
\citep{2012ApJ...755..155S,2012ApJ...750...58B,2012ApJ...759..132A,
2013arXiv1309.2313A,2012arXiv1212.0349S,2013arXiv1309.2313A}
and magnetohydrodynamic \citep{2013MNRAS.433.2165S} simulations,
showing the breakup of the cloud close to periastron and the infall of
cloud material towards the central BH.

For this scenario to be acceptable, one needs a valid
mechanism for the formation of the cloud, which will in principle be
disrupted in its first periastron passage. For example, the possible creation of
the cloud through the tidal disruption of the outer envelope of a
giant star has been discussed by \cite{2014arXiv1401.2990G}.

Scenario $b$ (see above) was modeled analytically by
\citet{scoville}. These authors show that a young, low mass
star with a mass loss rate $\sim 10^{-7}$~M$_\odot$~yr$^{-1}$ and a
terminal wind velocity $\sim 100$~km~s$^{-1}$ would
produce a cometary wind bubble with observational properties similar
to the G2 cloud. \cite{2013ApJ...776...13B} present isothermal, axisymmetric
simulations of this scenario, under the assumption of a
radial orbit. This is a reasonable approximation for the initial
approach of the cloud to the BH (as the orbit of G2 is highly
elliptical), but is inappropriate for modeling the periastron
passage.

This ``cometary wind bubble'' scenario for the G2 cloud avoids the
issue of the cloud origin. In each periastron passage, the wind
bubble will be disrupted. Then, at long
enough timescales after the periastron passage the wind bubble will be
regenerated by the continuing stellar wind pushing against the lower
density, outer environment.

In the present paper, we discuss 3D simulations of the
time-evolution of a stellar wind bow shock, following the passage through
the periastron of the G2 cloud orbit. The simulations include a model
for the coronal environment around the BH
\citep{2003ApJ...598..301Y},
the gravity force of the BH, and a parametrized
cooling function. The differences between our simulations and the ones
of \cite{2013ApJ...776...13B} is that our simulations are 3D (instead of
axisymmetric) and include a cooling function (instead of
using an isothermal approximation).


\section{Numerical Method and Setup}

\begin{figure*}
\centering
\includegraphics[width=\textwidth]{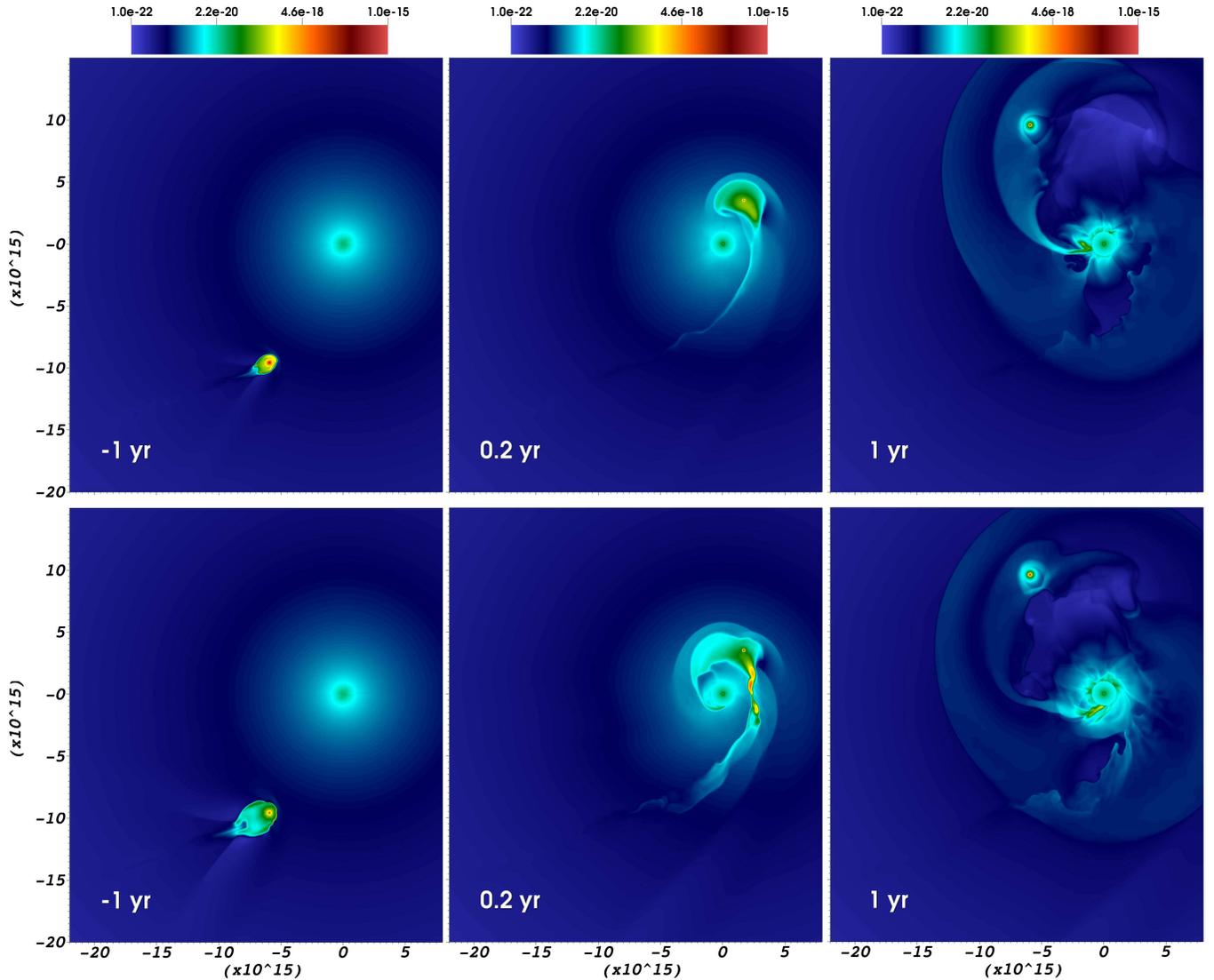}
\caption{Density stratification for the $v_{\rm w} = 50$ km s$^{-1}$ (\emph{top}) and
$v_{\rm w} = 200$ km s$^{-1}$ (\emph{bottom}) models (starting 3 yrs before periastron), for times 1 yr before periastron (left panels), and 0.2 and 1 yr after periastron (central and right panels respectively). The plot corresponds to a cut of the three-dimensional box along the $z=0$ plane. The BH is located at the origin of the coordinate system.}
\label{fig1}
\end{figure*}

\begin{figure}
\centering
\includegraphics[width=0.5\textwidth]{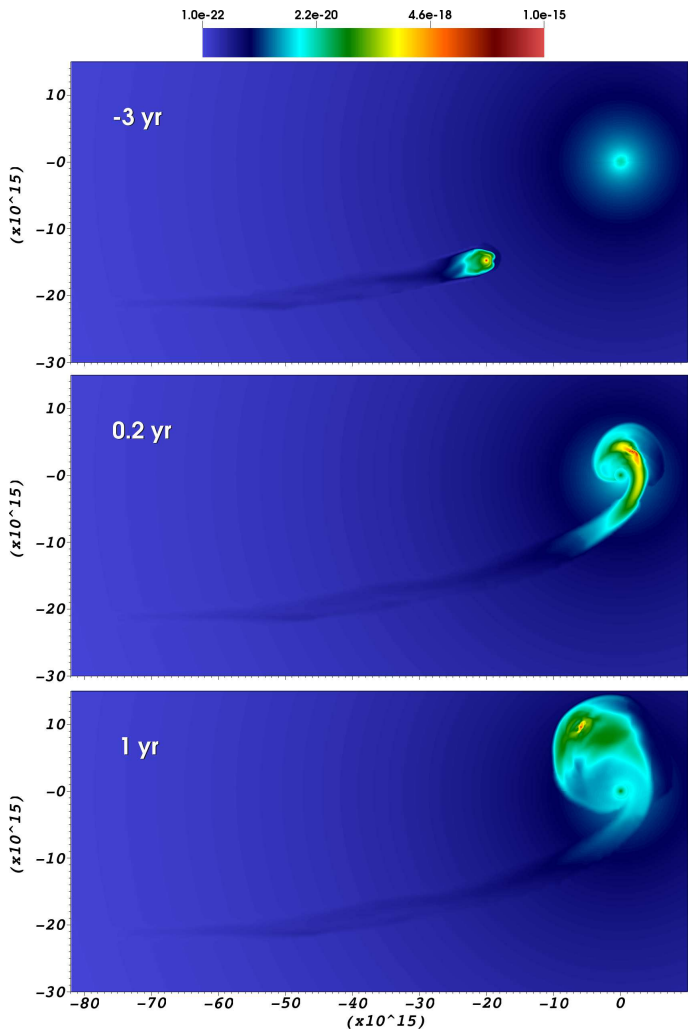}
\caption{The same as in Figure \ref{fig1}, but for a simulation with $v_{\rm w} = 50$ km s$^{-1}$ 
starting 19 yrs before periastron.}
\label{fig2}
\end{figure}

\begin{figure}
\centering
\includegraphics[width=0.5\textwidth]{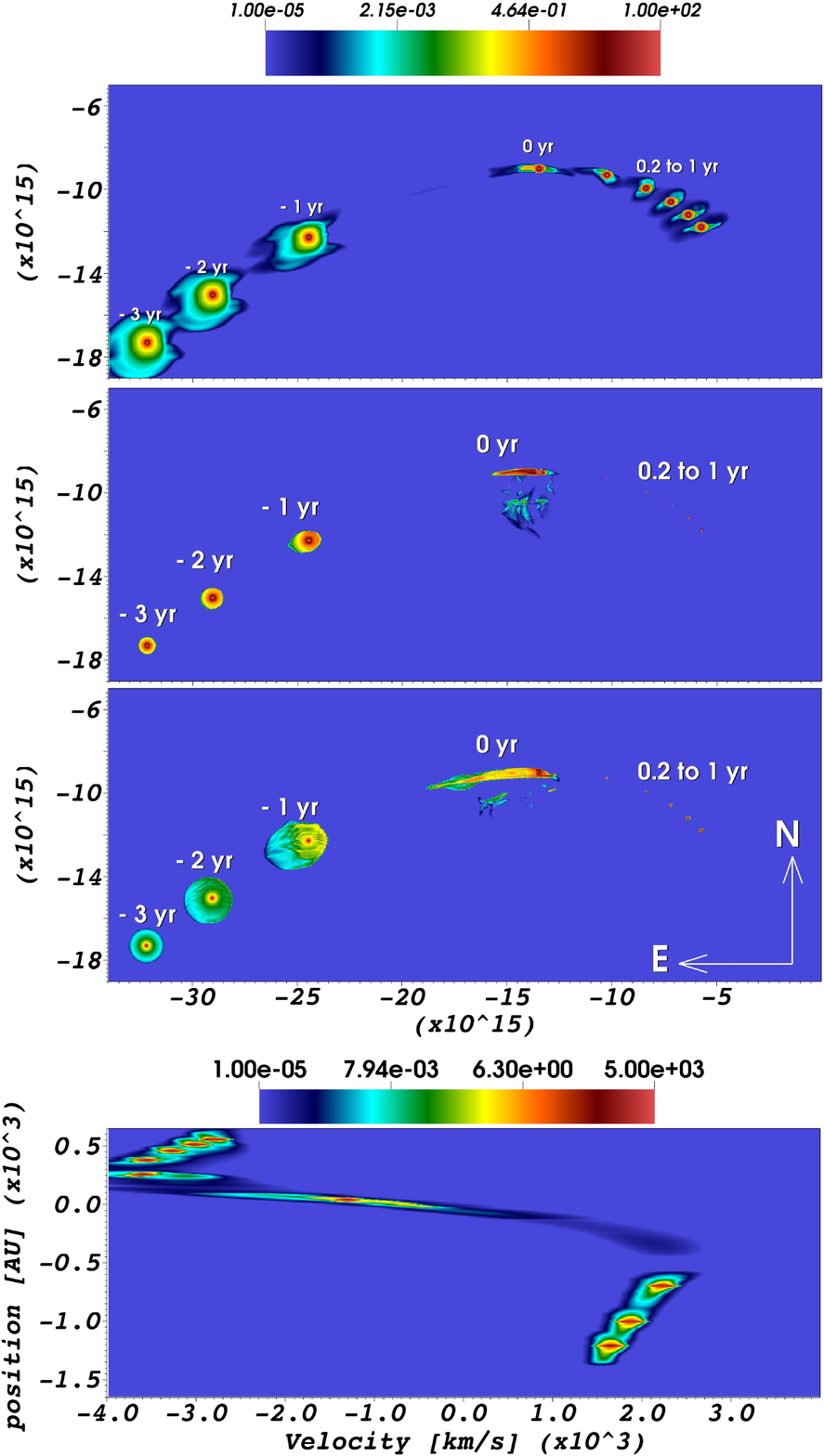}
\caption{\emph{Top panels}: Br$\gamma$ emission map (in units of [erg s$^{-1}$ cm$^{-2}$ sterad$^{-1}$]), 
projected on the plane of the sky
for the (from top to bottom) $v_{\rm w} = 50$ km s$^{-1}$ (20 and 4 yrs) and
$v_{\rm w} = 200$ km s$^{-1}$ models.
\emph{Bottom panel}: Position-Velocity diagram 
for the $v_{\rm w} = 50$ km s$^{-1}$, 20 yrs model.
The figure includes the emission occurring at different times, with the labels in the top panels indicating the emission time with respect to the periastron passage.
}
\label{fig3}
\end{figure}

We solve the hydrodynamic (HD)
equations using the adaptive mesh refinement code \emph{Mezcal}  \citep{decolle12},
extensively used to study astrophysical flows \citep[e.g.,][]{decolle06, decolle08, decolle12b}.
Our implementation of the HD equations includes the BH's gravitational force, 
with a $M_{\rm BH} = 4.31\times$10$^6$~M$_\sun$ BH mass \citep[e.g,][]{2008ApJ...689.1044G}, 
and a parametrized non-equilibrium cooling function \citep{1995MNRAS.275..557B}. 
We also take into account the effect of UV photons emitted by the massive stars present in 
the BH environment, by adding a heating term, which has the only effect of
keeping the gas temperature above 10$^4$~K. We assume that the H in the wind 
is fully ionized.

The exact stratification of the BH environment is uncertain.
In this paper, we take a density and pressure radial dependence 
of the form $\rho_a=9.5\times10^{-22} (10^{16} \mathrm{cm}/R)$~g\;cm$^{-3}$ and 
$p_a=G M_{\rm BH}\rho /R$ \citep{2012ApJ...750...58B,2012ApJ...759..132A,2013ApJ...776...13B,
2012ApJ...755..155S}, being $R$ the distance from the BH.
This stratified medium is convectively unstable. 
To avoid the development of the instability, 
\citet{2012ApJ...755..155S}, 
\citet{2013ApJ...776...13B}, and
\citet{2012ApJ...750...58B}
reset at every timestep the value of the BH
environment density to the unperturbed, hydrostatic value. 
This is the same strategy used in one of our simulations
(integrated for 20 yrs, see below).
We also run two simulations integrated during 4 yrs and without resetting
the BH environment, i.e. properly computing the bow shock
dynamics.

We assume that the G2 cloud is originated by the stellar wind (with a mass-loss rate 
$\dot{M_{\rm w}} = 10^{-7}$~M$_\sun$\;yr$^{-1}$) ejected by a young star,
which is moving in a very eccentric orbit ($e=0.966$) with a period of 198~yr 
\citep{2013ApJ...763...78G} around the super-massive BH.
The interaction of the wind with the ambient medium produces a double shock
structure, with an internal shock where the free-wind 
gas is decelerated, and an external shock where the ambient medium 
is accelerated. The distance from the double shock structure
to the center of the star can be estimated from the ram pressure balance
between the wind and the ambient medium (in the reference system of the 
star):
\begin{equation}
 r_{\rm sh} \approx \sqrt{\frac{\dot{M_{\rm w}} v_{\rm w}}{4\pi\rho_a v^2_{\rm a}}}
 \label{eq:rsh}
\end{equation}
where $v_w$ is the terminal velocity of the stellar wind,
$\rho_a$ is the (position dependent) environmental density and
$v_a$ is the orbital velocity of the stellar source of the G2 cloud.
The stellar wind is imposed in the code by rewriting at every timestep the 
values of the density and velocity inside a sphere with radius $0.2\times r_{\rm sh}$ 
(centered on the star) with the free-wind values.

To study the effect of changing the wind velocity and of injecting the
wind during different periods, we run three models: a high velocity 
model with a $v_{\rm w} =
200$~km\;s$^{-1}$\ wind velocity integrated for 4 yrs, starting 3 yrs before
periastron, and two low velocity models  with $v_{\rm w} = 50$~km\;s$^{-1}$,
and durations of 4 (starting 3 yrs before periastron) and 20 yrs
 (starting 19 yrs before periastron).
The density of the wind is calculated as $\rho = \dot{M}_{\rm w}/(4
\pi r^2 v_{\rm w})$ (where $r$ is the distance from the star). The pressure 
is chosen by fixing the temperature of the wind at 10$^4$~K, and using the
ideal gas equation of state. The exact value of the wind temperature is not 
important as the Mach number is $\gg 1$ in the free-wind region.

The simulations employ 3D Cartesian coordinates, with a volume 
of physical size extending from (-8.2$\times$10$^{16}$,-3$\times$10$^{16}$,0)~cm
 to (10$^{16}$,1.5$\times$10$^{16}$,10$^{16}$)~cm
 in the simulation lasting 20 yrs, and from
(-2.2$\times$10$^{16}$,-2$\times$10$^{16}$,0)~cm
 to (8$\times$10$^{15}$,1.5$\times$10$^{16}$,5$\times$10$^{15}$)~cm
in the simulations lasting 4 yrs
along the $x$-, $y$- and $z$-axes, respectively, sampled by using 
(184,90,20) and (96,112,16) cells at the coarsest level of refinement respectively. 
Reflective boundary conditions 
are used on the $z=0$ (orbital) plane.
The maximum number of levels of refinement is chosen by requiring
a minimum of 20 cells in $r_{\rm sh}$ (and is therefore changing with time).
As the star approaches periastron, $r_{\rm sh}$ (equation \ref{eq:rsh}) becomes increasingly 
small and a maximum of 8 levels of refinement are necessary
to properly resolve the free-wind region in the simulations lasting 4 yrs,
corresponding to a maximum resolution of 2.4$\times 10^{12}$~cm.
To limit the computational time, in the simulation lasting 20 yrs
$r_{\rm sh}$ is set as $r_{\rm sh} = \max (r_{\rm sh}, 10^{14}~{\rm cm})$.
The BH is located at the 
center of the reference system employed in the simulations, and the star moves 
in the $z=0$ plane.


\begin{figure}
\centering
\includegraphics[width=.5\textwidth,angle=0]{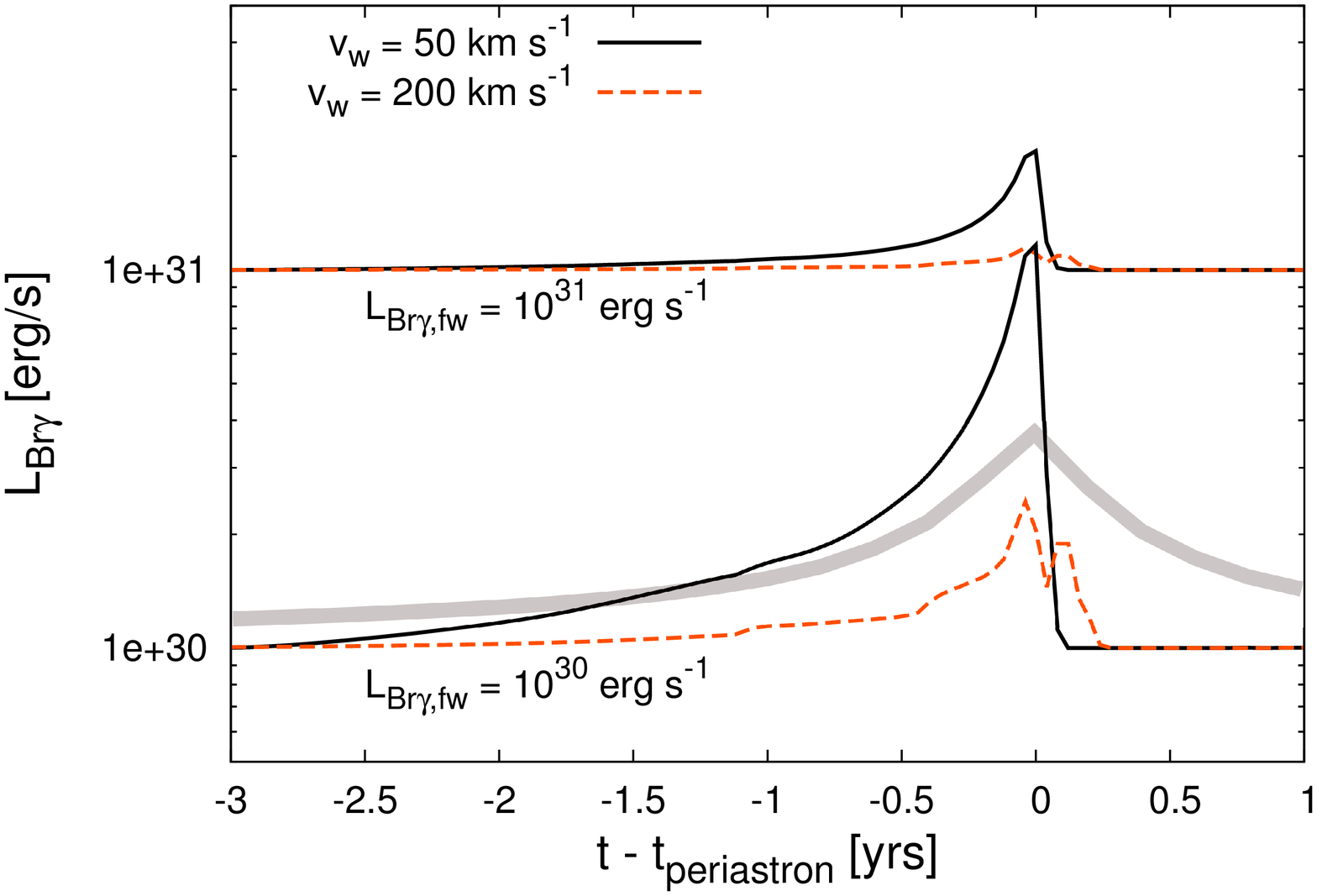}
\caption{
Time evolution of the integrated Br$\gamma$ luminosity, for the $v_{\rm w} = 50$ km s$^{-1}$, 4 yrs (black, solid line) and 20 yrs (thick, grey line) models, and $v_{\rm w} = 200$ km s$^{-1}$ (orange, dashed lines) model. The Br$\gamma$ luminosity is computed by integrating the flux emitted by the shocked wind and by adding a fixed luminosity of
$10^{30}$ erg s$^{-1}$ (bottom curves) and $10^{31}$ erg s$^{-1}$ (upper curves) to simulate
the emission coming from the (unresolved in the simulaton) inner region of the unshocked wind.}
\label{fig4}
\end{figure}

\section{Results}

Figures \ref{fig1} and \ref{fig2} show density slices on the orbital plane for integration
times $t=-1/-3$~yr (i.e., 1/3 years before periastron), 0.2 yr and 1 yr (after periastron) for
the $v_w=50$ and 200~km~s$^{-1}$ simulations:
\begin{itemize}
\item in the $t=-1$~yr frames
we see that the $v_w=200$~km~s$^{-1}$ model has a
cometary wind bubble $\sim 2$ times the size of the one of the
$v_w=50$~km~s$^{-1}$ simulation (as expected from equation
\ref{eq:rsh}),
\item in the $t=0.2$~yr time frames (central column of Figure  \ref{fig3}), the two
simulations show that the wind bubble has become a
broken filament, with a long, trailing tail and a shorter region
extending ahead of the position of the stellar wind source,
\item in the $t=1$~yr frames, the region around the BH is
  filled with a complex distribution of accreting, clumpy
  structures. The stellar wind is beginning to reform a cometary
  bubble (seen in the top region of the $v_w=50$ and 200 km s$^{-1}$
  simulations). Also, we see filamentary density structures connecting
  the regenerated wind bubbles to the region close to the BH,
\item Figure \ref{fig2} shows the effect obtained by
considering a wind injected during a much longer timescale.
Wind material accumulates around the central star, while a low 
density, extended tail forms behind the cometary wind bubble
which is also more extended in volume before periastron.
\end{itemize}

Figure  \ref{fig3} shows a composition of the predicted Br$\gamma$ intensity
maps for different evolutionary times and a position-velocity diagram
for the 20 yrs, $v_w=50$~km~s$^{-1}$ simulation.
The Br$\gamma$ emission coefficient is obtained by solving a 15 energy
level recombination cascade problem with the parameters of 
\citet{1970MNRAS.148..417B}. From a fit to the temperature dependence of the
Br$\gamma$ emission we obtain:
\begin{equation}
  L_{Br\gamma} = 9.55\times 10^{-29} \frac{n_e n_{\rm HII}}{\left(T/T_0\right)^{0.75}}  
   \left( 1-e^{-\left(\frac{T}{T_0}\right)^{-0.74}}\right) {\rm erg \; s^{-1}}\;,
\end{equation}
which reproduces the recombination cascade calculation within 3\%\ in the
$T=10^3\to 10^6$~K temperature range.

These intensity maps correspond to
the orientation of the G2 cloud orbit given by
\cite{2013ApJ...773L..13P}. These maps show the evolution of the cloud
from a cometary shape (for times $t<-1$~yr) to an elongated filament
(at $t\sim 0$) and finally to a more compact, centrally peaked
emission structure (in the months following periastron passage).
A relatively long time, of order $10^{15}$cm/$v_w$ ($\sim 6$~yr for
the $v_w=50$~km~s$^{-1}$ model) will be necessary for the cloud to
regain the size that it had at $t=-1$~yr before periastron.
The $v_w=50$~km~s$^{-1}$ model presents less 
extended emission at all evolutionary times.
The position-velocity diagram (Figure \ref{fig3}) shows 
an elongated emission extending a few hundred 
km s$^{-1}$ in the velocity channel, qualitatively similar to the 
observed one (e.g. \citealt{2013ApJ...763...78G}), becoming less
extended spatially after periastron passage
and much more extended in velocity space during and immediately
after periastron passage.

Figure \ref{fig4} shows the spatially integrated Br$\gamma$
luminosity as a function of time, computed by adding the contribution
of the cells located at a distance $R \geq R_{\rm sh}$
from the center of the star, i.e., the shocked wind, and a
fixed contribution of 10$^{30}$-10$^{31}$~erg~s$^{-1}$ 
assumed to come from the inner-region of the free-wind, and corresponding
to a recombination rate of 10$^{42}$-10$^{43}$~s$^{-1}$
 respectively.
The shocked wind component of the $v_w=50$~km~s$^{-1}$ model 
shows a larger Br$\gamma$ luminosity
(by a factor $\approx 10$) with respect to the
the $v_w=200$~km~s$^{-1}$ model.
Both models present a slow rise during the months/years preceding 
periastron passage, which is not present when the constant free-wind flux
dominates the total luminosity, and a fast decrease during the first month
after periastron. 

\begin{figure}
\centering
\includegraphics[width=.3\textwidth,angle=-90]{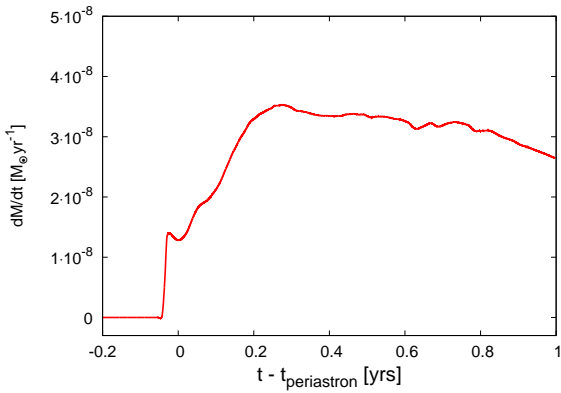}
\caption{Accretion mass rate on the BH (simulated as a free boundary with radius R$=10^{15}$~cm) as a function of time for the $v_{\rm w} = 50$ km s$^{-1}$, 20 yrs simulation.}
\label{fig5}
\end{figure}

Finally, in Figure  \ref{fig5} we show the mass accretion rate $\dot{M}_{\rm BH}$ 
for the  20 yrs, $v_w=50$~km~s$^{-1}$ model, computed by tracking the mass density flux 
across a $R=10^{15}$~cm sphere centered on the BH. $\dot{M}_{\rm BH}$
shows a quick increase during periastron passage, followed by a slow drop of the mass 
accretion rate after $\approx 0.2$~yr, with values of $\dot{M}_{\rm BH}$ of $\sim 3\times 10^{-8}$~M$_\sun$~yr$^{-1}$.
We can therefore expect that the mass accretion rate will remain large (above 
$10^{-8}$~M$_\sun$~yr$^{-1}$) for several years after periastron passage.


\section{Discussion}

Our 3D simulations of the interaction 
between a stellar wind and the Sgr A* environment show a 
complex dynamical behavior. 
As the star moves towards periastron, a weak forward shock
(with Mach number $\lesssim 2$) is formed.
Because of the rapid increase in the orbital velocity, the shocked 
region is unable to adjust to the equilibrium 
configuration expected from ram-pressure balance 
(equation \ref{eq:rsh}). As a consequence, the size of the interaction
region is larger (smaller) than predicted by equation \ref{eq:rsh} 
before (after) periastron passage. 
At periastron, strong tidal forces and the development of instabilities
break the wind bubble, forming an elongated structure with small, 
dense condensations.

These dense knots accrete onto the central region with a
 $\dot{M}_{\rm accr}$ of $\approx 3 \times
   10^{-8}$~M$_\sun$~yr$^{-1}$ accretion rate
(see Figure \ref{fig5}). These values are comparable to those obtained
in 3D simulations of the isolated clump scenario \citep{2012ApJ...759..132A},
and are also of the same order as the accretion rate on 
the BH inferred from the X-ray 
Sgr A* luminosity ($\dot{M}_{\rm BH}\sim 10^{-9}$-10$^{-7}$~M$_{\sun}$
yr$^{-1}$, see e.g. \citealt{2000ApJ...534L.173A}).
On the other hand, the spherical mass accretion rate 
\citep{1952MNRAS.112..195B}, computed at 
the distance from the BH where our value of $\dot{M}_{\rm acc}$ is calculated (R = 10$^{15}$~cm), 
is $\dot{M}_B \sim 4 \times 10^{-6}$~M$_\sun$~yr$^{-1}$ (e.g. \citealt{2003ApJ...598..301Y}).
As the streams of material accreting from G2 onto the inner region of the BH
(where the X-ray emission is thought to be generated) depends on
not fully understood physical processes (see, e.g., the discussion of \citealt{2012ApJ...759..132A}), it is 
possible that the small $\dot{M}_{\rm accr}$ obtained from the simulations
($\lesssim 0.02 \times \dot{M}_B$) could produce X-ray flares during the upcoming years.

No variations in the emission at radio wavelengths
(associated with relativistic electrons accelerated by the G2 forward shock) 
have yet been detected
\citep[e.g.,][]{2013ATel.5159....1B, 2014Atel.5727.1C}.
In the scenario of G2 as an isolated ISM cloud, \citet{2012ApJ...757L..20N}
(see also \citealt{2013MNRAS.432..478S})
predicted detectable radio emission with a flux $\sim$ ten times the Sgr A* quiescent flux, while, more recently,
 \citet{2014ApJ...783...31S} obtained a 30 times lower radio flux
 (for the same scenario)
 by assuming a lower size for the radio emitting region.

In the context of a stellar wind model, \citet{2013MNRAS.436.1955C}
 have shown that the radio emission decreases with the forward shock size
as the star approaches periastron. 
Our simulations confirm that the 
forward shock size drops quickly when the star approaches 
periastron indicating that it is 
unlikely to produce detectable radio emission. 
However,  a detailed calculation based on the results
of numerical simulations is necessary to make a quantitative prediction of the 
radio emission, as a simple analytical 
approach fails to describe the elongated, broken shock structure observed during
the periastron passage.

Observations of the G2 cloud with the \emph{Integral Field 
Spectrograph} on the VLT, SINFONI, show a compact Br$\gamma$ emission 
from the head of G2 with a luminosity of $\sim 8 \times 10^{30}$~erg s$^{-1}$, 
and an extended tail with approximately 
the same luminosity \citep{2013ApJ...776...13B}. 
Our low-velocity ($v_w=50$~km~s$^{-1}$) simulation does produce
a compact emission and an extended tail with comparable Br$\gamma$
luminosities (see Figure \ref{fig3}). 

The Br$\gamma$ 
luminosity of the G2 cloud has remained nearly constant (within the statistical errors)
during the last $\sim 9$~ yrs \citep{2013ApJ...763...78G}, while 
in both the stellar wind and the isolated clump
one would expect an increase in the flux as ionization of neutral gas 
should become more efficient approaching periastron \citep{scoville}.
In our simulations, both the gas in the free-wind region and the 
shocked stellar wind material contribute to the Br$\gamma$ emission.
The emission from the shocked ambient medium gas is negligible as it is
too hot and rarefied to emit efficiently at infrared wavelengths. 

Figure \ref{fig4} shows that the emission from the shocked stellar wind 
increases approaching periastron (in agreement with 
\citealt{2013ApJ...776...13B}). The interpretation of the observation
remains therefore puzzling. To explain it, we assume the presence of
a strong ambient photoionizing radiative field (resulting in a
strong Br$\gamma$ emission from the free wind region), larger
than the variability obtained from the simulations in the total Br$\gamma$ luminosity.
A solution to this problem, requiring detailed modeling of the BH environment
including radiation transfer, is beyond the scope of this paper.

Our simulations illustrate the fact that (in the stellar wind bow shock
scenario) once the G2 cloud passes through periastron and the
tail of dense, ionized gas moves toward
the central BH, the Br$\gamma$ emission becomes 
less extended and more peaked on the position of the star (see Figure \ref{fig3}). 

A caveat in the interpretation of our 3D simulations is that they do
not properly resolve the dense, low filling factor gas seen in
the axisymmetric simulations of \cite{2013ApJ...776...13B}. Therefore,
one might expect that higher Br$\gamma$ luminosities might be
obtained in simulations with higher resolutions. This can be seen e.g.
comparing the Br$\gamma$ emission of the low velocity models in Figure
\ref{fig4}.

Recent observations by Ghez et al. (2014) show that G2 is still 
intact at periastron passage, indicating that a stellar component is 
probably present within the G2 cloud.
Clearly, the issue of whether or not the G2 cloud is fed by a central,
stellar source will be elucidated by observations over the next few
months. If there is indeed a stellar component to G2, a compact Br$\gamma$
source should survive periastron passage and continue its motion along
the G2 orbit. This compact source should become more extended over the
next few years, as the stellar wind bow shock develops again.

The simulations presented in this paper are a first attempt at
modelling the destruction of a stellar wind bubble on passage through
the G2 periastron. In about a $\sim 1$ yr timescale, once
observations of the periastron passage clarify the origin of 
the G2 cloud, it might become worthwhile to pursue more detailed models, e.g.,
including the rotation of the BH environment
\citep{2013arXiv1309.2313A} and/or the transfer of ionizing radiation.


\acknowledgments 
We aknowledge P. Crumley, D. Gonz\'alez-Casanova, J. Guillochon, L.F. Rodriguez for helpful discussions. 
This research was supported by the DGAPA-PAPIIT-UNAM grants IA101413-2, IG100214, IN105312, IN106212, and the CONACyT grants 101356, 101975, 165584, 167611 and 167625.
FFC-T acknowledges ICN-UNAM for supporting his postdoctoral stay. 


\end{document}